# Evidence of Gate-Tunable Mott Insulator in Trilayer Graphene-Boron Nitride Moiré Superlattice

Guorui Chen[1,2,6], Lili Jiang[1], Shuang Wu[5], Bosai Lv[4,6], Hongyuan Li[4,6], Bheema Lingam Chittari[8], Kenji Watanabe[7], Takashi Taniguchi[7], Zhiwen Shi[4,6], Jeil Jung[8], Yuanbo Zhang[2,6,9] *, Feng Wang[1,3,10] *

[1]Department of Physics, University of California at Berkeley, Berkeley, CA, USA.
[2]State Key Laboratory of Surface Physics and Department of Physics, Fudan University, Shanghai 200433, China.
[3]Materials Science Division, Lawrence Berkeley National Laboratory, Berkeley, CA, USA.
[4]Key Laboratory of Artificial Structures and Quantum Control (Ministry of Education), School of Physics and Astronomy, Shanghai Jiao Tong University, Shanghai, China.
[5]Department of Electrical Engineering and Computer Sciences, University of California, Berkeley, CA 94720, USA.
[6]Collaborative Innovation Center of Advanced Microstructures, Nanjing, China.
[7]National Institute for Materials Science, 1-1 Namiki, Tsukuba, 305-0044, Japan.
[8]Department of Physics, University of Seoul, Seoul 02504, Korea.
[9]Institute for Nanoelectronic Devices and Quantum Computing, Fudan University, Shanghai 200433, China
[10]Kavli Energy NanoSciences Institute at the University of California, Berkeley and the Lawrence Berkeley National Laboratory, Berkeley, CA, USA.

*e-mail: fengwang76@berkeley.edu; zhyb@fudan.edu.cn;

**Mott insulator plays a central role in strongly correlated physics, where the repulsive Coulomb interaction dominates over the electron kinetic energy and leads to insulating states with one electron occupying each unit cell.[1,2] Doped Mott insulator is often described by the Hubbard model[3], which can give rise to other correlated phenomena such as unusual magnetism and even high temperature superconductivity[2,4]. A tunable Mott insulator, where the competition between the Coulomb interaction and the kinetic energy can be varied *in situ*, can provide an invaluable model system for the study of Mott physics. Here we report the realization of such a tunable Mott insulator in the ABC trilayer graphene (TLG) and hexagonal boron nitride (hBN) heterostructure with a moiré superlattice. Unlike massless Dirac electrons in monolayer graphene[5,6], electrons in pristine ABC TLG are characterized by quartic energy dispersion and large effective mass[7-11] that are conducive for strongly correlated phenomena. The moiré superlattice[12-18] in TLG/hBN heterostructures leads to narrow electronic minibands that are gate-tunable. Each filled miniband contains 4 electrons in one moiré lattice site due to the spin and valley degeneracy of graphene. The Mott insulator states emerge at 1/4 and 1/2 fillings, corresponding to one electron and two electrons per site, respectively. Moreover, the Mott states in the ABC TLG/hBN heterostructure exhibit unprecedented tunability: the Mott gap can be modulated *in situ* by a vertical electrical field, and at the mean time the electron doping can be gate-tuned to fill the band from one Mott insulating state to another. Our observation of a tunable Mott insulator opens up exciting opportunities to explore novel strongly correlated phenomena in two-dimensional moiré superlattice heterostructures.**

Atomically thin two-dimensional (2D) materials with wide-ranging properties can be grown separately and then stacked together to form a new class of materials – van der Waals-bonded heterostructure, in which each layer can be engineered separately.[19] In addition, convenient control of individual layered materials may be achieved through electrostatic gating and inter-layer coupling. Such van der Waals heterostructures offer the possibility to design new material systems that were simply not possible in the past. Many fascinating physics have already been observed in different van der Waals heterostructures, such as mini-Dirac cones and the Hofstadter's butterfly pattern in graphene/hBN heterostructures[12-15,18,19], direct to indirect bandgap transitions in bilayer MoS$_2$[22,23], and interlayer exciton states in MoSe$_2$/WSe$_2$ heterostructures[24,25]. So far such new phenomena in van der Waals heterostructures are limited to weakly interacting electrons. Novel strongly correlated physics can also emerge in suitably designed van der Waals heterostructures where the Coulomb potential dominates over the kinetic energy of band electrons. A quintessential example of correlated physics is the Mott insulator, which features an insulating state in partially filled electronic band, usually at the density of one electron per lattice site. Previously a one-dimensional Mott insulator was observed in suspended single-walled carbon nanotubes due to enhanced Coulomb interactions at lower dimensions. Here we design an ABC-stacked TLG and hBN heterostructure with a large moiré superlattice to realize for the first time a tunable 2D Mott insulator, where the Mott gap can be modulated *in situ* by a vertical electrical field. In addition, the electron doping can be continuously gate-tuned to fill the electronic band from one Mott insulating state to another, which is impossible to achieve for Mott insulator based on bulk natural crystals.

Graphene/hBN heterostructures offer a powerful platform to explore novel quantum phenomena due to the excellent sample quality and tunable physical property [26] of graphene and the existence of a periodic moiré superlattice[13-15]. In monolayer graphene, the massless electrons feature a large kinetic energy, resulting in moiré minibands with rather wide energy bandwidths. Consequently, electron-electron interaction effect is negligible for monolayer graphene/hBN superlattice. In contrast, ABC-stacked TLG exhibits a very flat quartic band at the K point in the first Brillouin zone[7-11]. Such flat band gives rise to strong electron correlation and spontaneously symmetry breaking state in suspended ABC TLG[9,27]. The addition of moiré superlattice with a $L_M$ = 15 nm period in zero-twisting ABC TLG/hBN heterostructure creates narrow minibands that are well separated from each other. The lowest minibands have a bandwidth around 10 meV, which is significantly smaller than the on-site Coulomb repulsion energy of $\frac{e^2}{4\pi\varepsilon_0\varepsilon L_M} \sim 25$ meV if we use the hBN dielectric constant of $\varepsilon = 4$. In this letter we report that the dominating Coulomb potential leads to prominent Mott insulating states with one hole per moiré superlattice site in the ABC TLG/hBN heterostructure (Fig. 1a), which corresponds to ¼ filling of the miniband due to the spin and valley degeneracy of electrons in graphene. Mott insulating states are also observed at ½ filling with two holes per superlattice site and at several other fraction fillings of the minibands. In addition, the band structure of the TLG can be controlled through electrostatic gating, leading to a field tunable Mott insulator in the heterostructure.

Few layer graphene samples are mechanically exfoliated onto Si wafer covered with 285 nm thick $SiO_2$, and the layer thickness is determined through optical contrast measurements. We use near-field infrared (IR) nanoscopy to identify ABC-stacked TLG because it allows direct imaging of ABC and ABA regions with nanometer spatial resolution[28]. Fig. 1b displays the topography image of an exfoliated graphene with monolayer, bilayer and trilayer segments on $SiO_2$/Si substrate using atomic force microscopy (AFM), and Fig. 1c shows the corresponding near-field nanoscopy image. The ABA and ABC regions of the TLG segment exhibit very different contrast in the near-field nanoscopy image due to their different electronic bandstructures and infrared responses[28]. We isolate the large ABC trilayer region by cutting the sample *in situ* with the tip of the AFM, and create the hBN/TLG/hBN heterostructure by stacking different layers with dry transfer method[26]. We identify the crystal orientation of TLG and hBN using the crystalline edges of the flakes and manually align the TLG lattice with the hBN flake during transfer process. The device is then etched into a Hall bar structure using standard e-beam lithography. The TLG is contacted through one-dimensional edge contacts with Cr/Pd/Au electrodes. We further deposit a metal top electrode to form a dual-gate devices where the TLG/hBN heterostructures can be gated by both the top metal electrode and the bottom silicon substrate. This dual-gate configuration allows us to control the carrier concentration and bandgap of TLG independently.[29] Figure 1d displays the optical image of a fabricated device, and Fig. 1e shows the schematic cross-sectional view of the device. We find that the ABC trilayers often convert into ABA trilayers after the fabrication process, and only a small fraction retain the original stacking order. We confirm the ABC stacking order of TLG after the fabrication process by characterizing its gate-dependent resistivity and Landau Level (LL) fan diagram in magnetic field. (See SI section I for details.)

The TLG/hBN heterostructures can form moiré superlattices with different lattice period. The moiré superlattice effect is strongest for zero-twisting graphene and hBN layers, which has the longest moiré period of $L_M$ = 15 nm. A key signature of the moiré superlattice is the new resistance peaks that emerge when the moiré minibands are completely filled. Such resistance peaks can arise from secondary Dirac points with monolayer graphene, or fully gapped insulating states with bilayer or ABC TLG. [13-16,26] Fig. 2a shows the gate-dependent four-probe resistivity ($\rho_{xx}$) at 1.4 Kelvin in a near-zero-twist ABC trilayer/hBN heterostructure, where the top and bottom hBN thicknesses are 38 nm and 40 nm, respectively. In this measurement, we fixed the bottom gate voltage at 0 V, and sweep the top gate voltage from -4.5 to 4.5 V. In addition to a large resistance peak at the charge neutral point (CNP) at $V_t^0 = -0.2$ V, two extra prominent resistance peaks appear symmetrically over the CNP at the gate voltages $V_t - V_t^0 = \pm 3.8$ V. They correspond to an electron and hole density of $n_{FFP}$ = 2.1×10$^{12}$ cm$^{-2}$ based on the capacitance model using a top hBN thickness of 38 nm and an hBN vertical dielectric constant of 4. The secondary resistance peaks are similar to those observed in monolayer and bilayer graphene/hBN moiré superlattices. They arise from fully filled moiré superlattice minibands, and we denote them as fully filled points (FFPs). The moiré superlattice period can be obtained by the relation $L_M = \sqrt{\frac{8}{\sqrt{3}n_{FFP}}} = 15$ nm. An independent and more accurate determination of the $n_{FFP}$ and the moiré wavelength through the Landau level fan diagram confirms these results. (See SI section II for details.) The emerging feature that distinguishes the ABC TLG/hBN heterostructure is the extra prominent resistance maxima in partially filled hole miniband, including a resistance peak close to ½ filling. This is in striking contrast to a typical band insulator (including the monolayer or bilayer graphene/hBN superlattice), where the resistance is close to a minimum value at ½ filling. Such resistance peak close to ½ filling implies strongly correlated electronic states in the ABC TLG/hBN heterostructure.

Further enhancement of the quantum correlation in electronic states can be achieved by suppressing the energy bandwidth of electronic bands. This can be realized in ABC TLG/hBN heterostructure through a vertical electrical field, which is known to induce a finite bandgap in ABC TLG. Experimentally, we control the vertical electrical field strength and the carrier doping in the ABC TLG/hBN heterostructure independently by varying both the top and bottom gate voltages[29]: the vertical displacement field across the TLG is set by $D = \frac{1}{2}(D_b + D_t)$ and the charge concentration is determined by $n = (D_b - D_t)/e$. Here $D_b = +\varepsilon_b(V_b - V_b^0)/d_t$ and $D_t = -\varepsilon_t(V_t - V_t^0)/d_t$, where $\varepsilon$ and $d$ are the dielectric constant and thickness of the dielectric layers, respectively, and $V_b^0$ and $V_t^0$ are effective offset voltages caused by environment-induced carrier doping. Fig. 2c shows a two-dimensional plot of the $\rho_{xx}$ value as a function of $V_t$ and $V_b$. The resistivity peaks at the CNP and FFPs persist at all displacement fields. In particular, the resistivity at the CNP increases monotonically with the vertical displacement field strength, as shown in Fig. 2d, indicating an increased band gap energy at the CNP. More strikingly, the resistivity peaks at partial filling become much sharper and more prominent at higher vertical displacement fields. Resistivity peaks can be clearly identified at 1/4 and 1/2 filling of the first hole miniband, and at 1/2 filling of the first electron miniband (dashed lines in Fig. 2b). The difference between the electron and hole minibands presumably arises from the electron-hole asymmetry present in ABC TLG.[7] Fig. 2b shows a horizontal line cut of Fig. 2c at $V_b$ = 20 V. The prominent resistance peaks at 1/4 and 1/2 fillings of the hole miniband are very sharp, and they are comparable, or even more insulating than the CNP and FFPs. The 1/4 and 1/2 filling states correspond to one electron and two electrons per superlattice unit cell, respectively, and this observation is the defining signature of a Mott insulator. Magneto-transport data further show that the resistance peaks at 1/4 and 1/2 filling states also feature zero Hall carrier density, further indicating the existence of a gapped insulator. (See SI section III for details.) Such Mott insulator in ABC TLG/hBN superlattice offers unprecedented control of the Mott states: the insulating states can be strongly modified in situ by changing the vertical electrical field, and the carrier doping can be gate-tuned through the whole Mott band so that we can change from one Mott insulating state to another Mott insulating state.

To estimate the Mott gap of the insulating states, we measured the temperature dependence of the transport behavior in the ABC TLG/hBN heterostructure. Fig. 3 shows a representative data set of resistivity versus temperature at $V_b$ = 20 V. The resistivity peaks at 1/4 and 1/2 fillings exhibit typical insulating behavior where the resistance increases with reduced temperature. The inset shows the ln$\rho_{xx}$-1/T plot for the ½ filling point, from which we can estimate an "effective transport gap" $\Delta_t$ of ~ 2meV.

Next we examine theoretically the competition between the moiré miniband bandwidth (W) and the Coulomb repulsion energy (U) in the ABC TLG/hBN heterostructure with an $L_M$ = 15 nm moiré superlattice. The single-particle bandstructure of the heterostructure is described by the Hamiltonian $H = H_{ABC} + V_M$, where $H_{ABC}$ is the TLG Hamiltonian under a weak vertical electrical field, and $V_M$ describes the effective potential acting on TLG from the moiré superlattice. The low-energy electronic structure of ABC-stacked TLG can be captured by an effective two-component Hamiltonian in the K valley that describes hopping between the A atom in the top graphene layer and the C atom in the bottom graphene layer[7,8],

$$H_{ABC} = \frac{v_0^3}{t_1^2}\begin{pmatrix} 0 & (\pi^+)^3 \\ \pi^3 & 0 \end{pmatrix} + \left(\frac{2v_0v_3p^2}{t_1} + t_2\right)\begin{pmatrix} 0 & 1 \\ 1 & 0 \end{pmatrix} + \left(\frac{2v_0v_4p^2}{t_1} - \Delta'\right)\begin{pmatrix} 1 & 0 \\ 0 & 1 \end{pmatrix} + \left(\frac{3v_0^2p^2}{t_1^2} - 1\right)\Delta''\begin{pmatrix} 1 & 0 \\ 0 & 1 \end{pmatrix} - \Delta\begin{pmatrix} 1 & 0 \\ 0 & -1 \end{pmatrix};$$

where $\pi = p_x + ip_y$, p is the electron momentum, 2Δ is the electron energy difference between the top and bottom layer due to the vertical electrical field $v_i \equiv \left(\frac{\sqrt{3}}{2}\right)at_i/\hbar$, a is the carbon-carbon lattice constant ~2.46 Å, $\Delta' \approx 0.0122$ eV, $\Delta'' \approx 0.0095$ eV, and $t_0, t_1, t_2, t_3, t_4$ are tight binding parameters in ABC TLG obtained from LDA *ab initio* calculations with values of 2.62 eV, 0.358 eV, -0.0083 eV, 0.293 eV and 0.144 eV, respectively. We consider that the encapsulated ABC TLG forms a near-zero-twisting moiré superlattice with the hBN on one side, for example, the bottom hBN film. The TLG/hBN interaction in the K valley can be approximated by a potential of the form $V_M^{A/B}(r) = 2\,C_{A/B}\,Re[e^{i\varphi_{A/B}}f(r)]\begin{pmatrix} 1 & 0 \\ 0 & 0 \end{pmatrix}$ acting at the low energy site of the effective ABC trilayer in contact with the substrate[17], where $f(r) = \sum_{j=1}^{6} e^{iq_j r}(1 + (-1)^j)/2$ and $\boldsymbol{q}_j$ are the six reciprocal lattice vectors of the triangular moiré superlattice with $|\boldsymbol{q}_j| = \boldsymbol{q}_M \equiv \frac{4\pi}{\sqrt{3}L_M}$. The hBN layer periodically modulates the potential in the bottom layer carbon atom whose magnitude and phase parameters are $C_A = -14.88$ meV, $\varphi_A = 50.19°$.[17,18] We solve the Hamiltonian numerically by direct diagonalization with a momentum cutoff at 5 $q_M$. Fig. 4a displays the energy dispersion of the two lowest electron and hole minibands in the ABC TLG/hBN heterostructure without an electrical field. The first hole miniband has a bandwidth of W ~ 20 meV. Upon applying an external vertical displacement field of 0.4 V/nm, we generate a potential difference between the top and bottom layer of TLG of ~ 20 meV [8]. The calculated miniband dispersion for $2\Delta = 20$ meV is displayed in Fig. 4b. We observe that the first hole miniband is strongly suppressed by the vertical field and has W ~ 13 meV. In particular, the hole miniband is well separated from other bands by over 10meV. The on-site Coulomb repulsion energy can be estimated by $U \sim \frac{e^2}{4\pi\varepsilon_0 \varepsilon L_M}$. For $L_M = 15$ nm and an hBN dielectric constant $\varepsilon = 4$, U is around 25meV, which is larger than the value of W. This dominating on-site Coulomb repulsion naturally leads to Mott insulator states in the isolated hole miniband when there are one or two holes per site, i.e. at ¼ and ½ filling of the band. The expected Mott gap ($\Delta_{Mott} \sim$ U-W) should be around 10 meV, which agrees qualitatively with our experimental observation. More information about the minibands electronic bandstructure and their evolution with the vertical electrical field are provided in SI section V.

The concept of engineering a moiré superlattice that exhibits correlated behavior by controlling the competition between the kinetic and potential energy is generally applicable to many atomically thin 2D heterostructures, including twisted semiconducting transition metal dichacogenide heterostructures that feature a reasonably large electron mass. Such tunable Mott insulating states in designed 2D heterostructures opens us completely new ways to explore fascinating Mott physics: The doped Mott insulator in 2D heterostructures can provide a unique tunable quantum simulator of the Hubbard model. The 2D heterostructures can also give rise to new correlated phenomena that are not present in conventional crystals, such as the correlations between the charge, spin, and valley degrees of freedom, and the interplay between the Mott insulator and quantum Hall states under high magnetic field.

**Supplementary Information**

Supplementary Information is available in the online version of the paper.

**Data Availability**

The data that support the findings of this study are available from the corresponding authors upon reasonable request.

**Acknowledgements**

We thank Chenhao Jin, Emma Regan, Xiaobo Lu, Yuwei Shan, Prof. Shiwei Wu and Prof. Guangyu Zhang for discussion and their help with sample preparation. The trilayer graphene sample fabrication and experimental study is supported by the Office of Naval research (award N00014-15-1-2651). The initial idea and proof of principle calculation of 2D flatband engineering was supported by an ARO MURI award (W911NF- 15-1-0447). J. J. acknowledges financial support from the Korean NRF through the grant NRF-2016R1A2B4010105, and assistance from B. L. Chittari for help with figure preparation. Y.Z. acknowledge financial support from National Key Research Program of China (2016YFA0300703), and NSF of China (grant nos. U1732274, 11527805 and 11425415). Z.S. is supported by the Program for Professor of Special Appointment (Eastern Scholar) at Shanghai Institutions of Higher Learning and the National Natural Science Foundation of China under Grant 11574204. Growth of hexagonal boron nitride crystals was supported by the Elemental Strategy Initiative conducted by the MEXT, Japan and JSPS KAKENHI Grant Numbers JP15K21722. Part of the sample fabrication was conducted at Fudan Nano-fabrication Lab.

**Author contributions**
F.W. and Y.Z. supervised the project. G.C. fabricated samples and performed transport measurements. G.C., L.J., S.W., B.L., H.L. and Z.S. prepared trilayer graphene and performed near-field infrared and AFM measurements. J. J. calculated the band structure. K.W. and T.T. grew hBN single crystals. G.C., Y.Z. and F.W. analyzed the data.

**Author Information**
The authors declare no competing financial interests. Correspondence and requests for materials should be addressed to F.W. (fengwang76@berkeley.edu) and Y.Z. (zhyb@fudan.edu.cn).

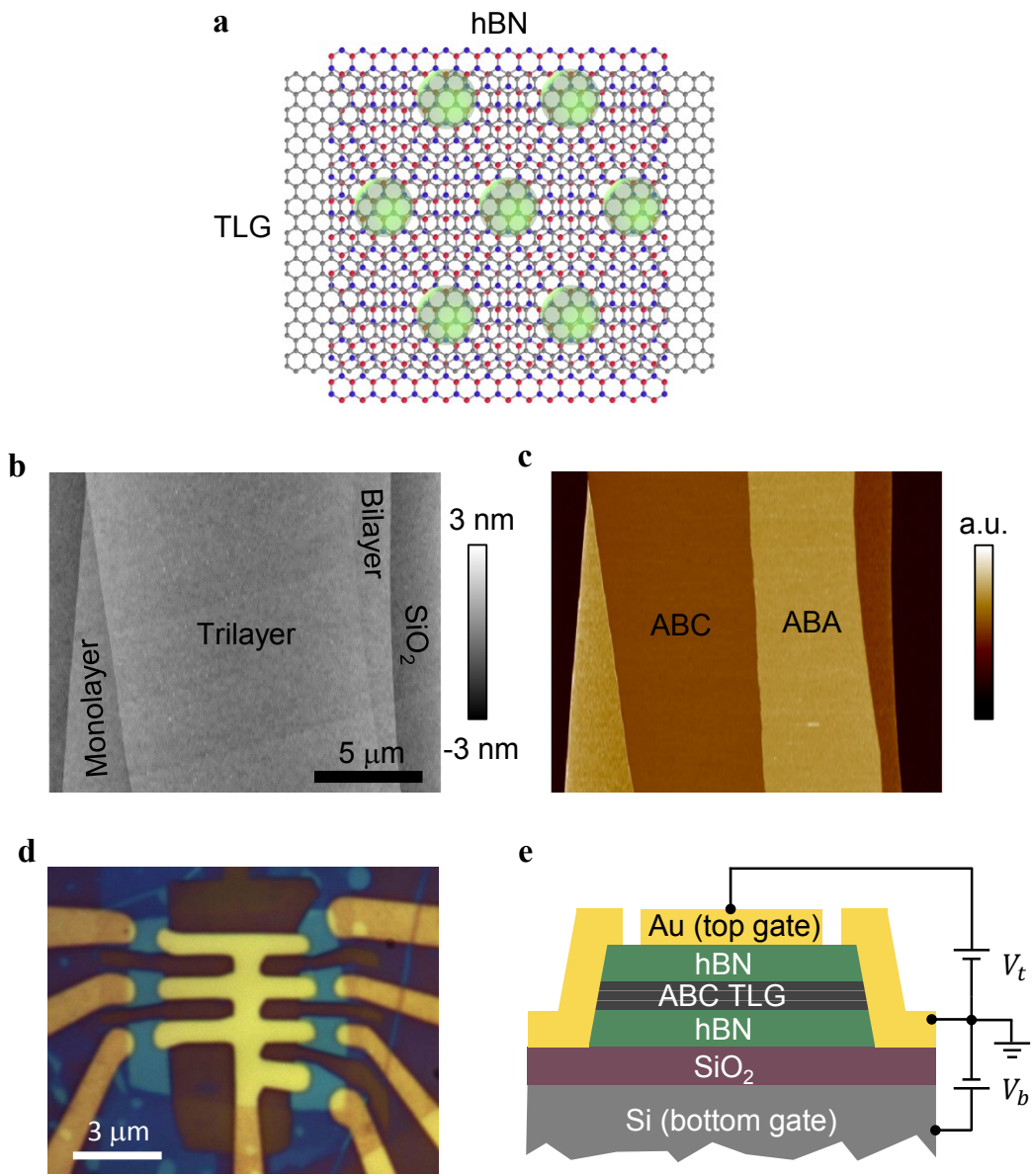

**Figure 1: ABC TLG/hBN Moiré superlattice and dual gate FET. a**, Schematic of ABC TLG/hBN Moiré superlattice. Only atoms of the top hBN layer and the bottom graphene layer are shown for image clarity. In the Mott insulating state at 1/4 filling, each electron (or hole) occupies one superlattice site and they are separated by the dominating Coulomb repulsion. **b**, AFM topography image of an exfoliated graphene with monolayer, bilayer and trilayer segments on $SiO_2$/Si substrate. **c**, The corresponding near-field infrared nanoscopy image in which a large ABC domain exists. **d**, Optical micrograph of a dual-gated ABC TLG encapsulated by hBN. The sample is etched into Hall bar for four-probe measurement and contacted by Cr/Pd/Au through the exposed edges. **e**, Schematic cross-sectional view of the device shown in **d**. The Au bar and doped Si served as top and bottom gate, respectively.

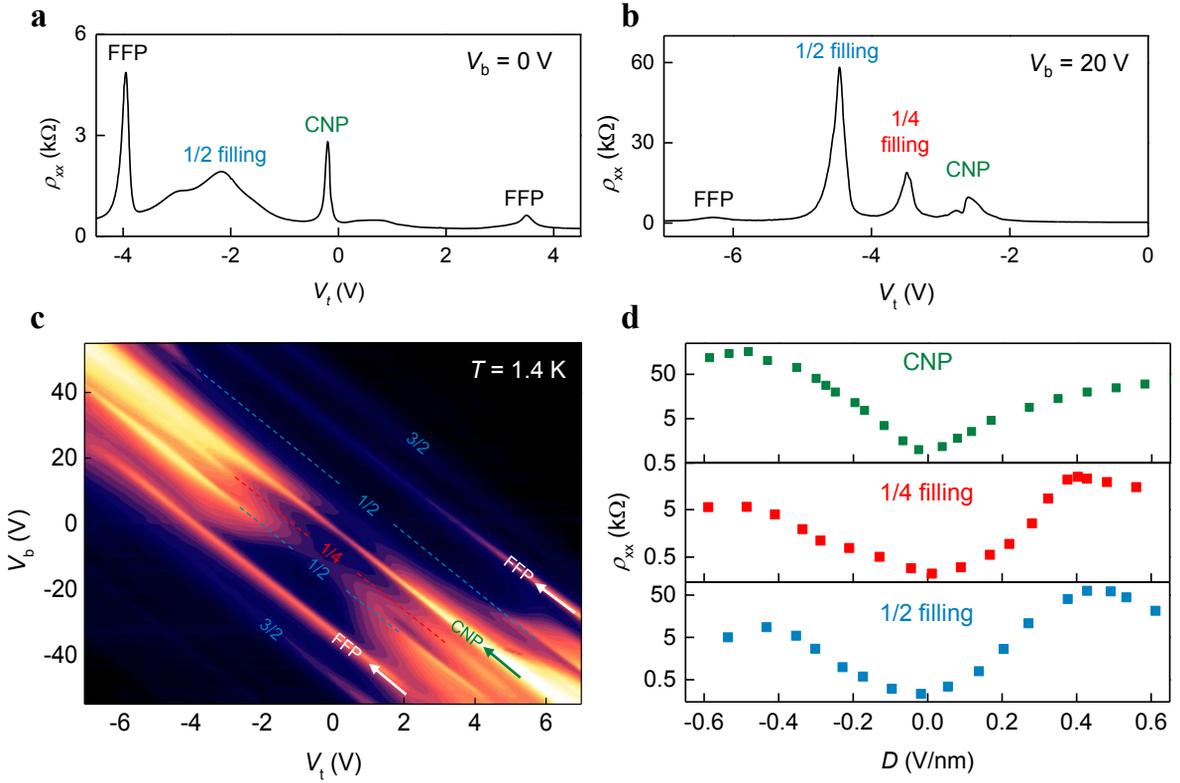

**Figure 2: Transport of gate tunable Mott state. a,b,** Top gate dependent resistivity of ABC TLG/hBN Moiré superlattice when $V_b$ = 0 V and 20 V, respectively. **c,** Color plot of resistance as a function of $V_t$ and $V_b$. The color-scale is from 10 Ω (dark) to 100 kΩ (bright) in log scale. The highlighted straight lines correspond to CNP, 1/4 filling, 1/2 fillings, FFPs and 3/2 fillings resistance peaks. **d,** Resistivity of CNP (green), 1/4 filling in the hole side (red), and 1/2 filling in the hole side (blue) tuned by electric displacement $D$. Insulating behavior at 1/4 and 1/2 fillings, correspond to one and two charge per lattice site, provides the defining signature of a Mott insulator.

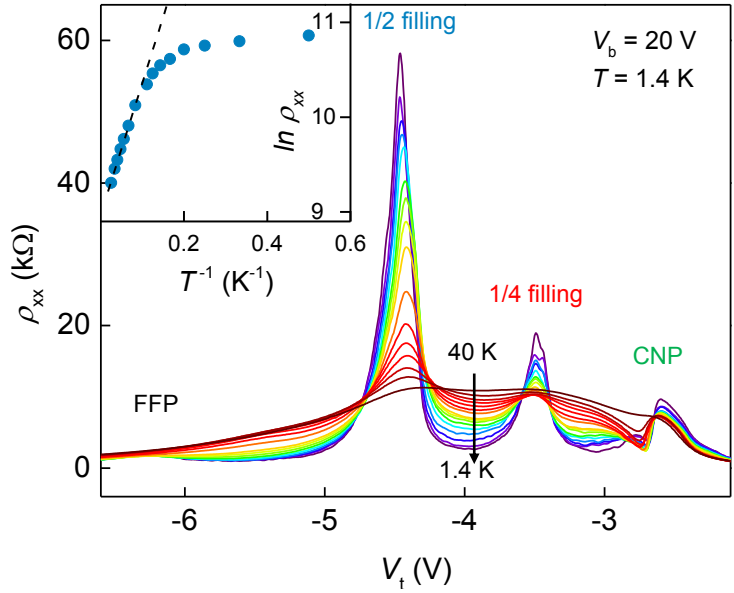

**Figure 3: Temperature dependent resistivity.** $\rho_{xx}$ at different top gate voltages and a fixed $V_b = 20$ V for temperatures ranging from 1.5 K to 40 K. The resistivity peaks at 1/2 filling and 1/4 filling exhibit typical insulating behavior where the resistance increases with reduced temperature. Inset: the corresponding $ln\rho_{xx} - 1/T$ plot at 1/2 filling, where the estimated transport Mott gap $\Delta_t = 2$ meV.

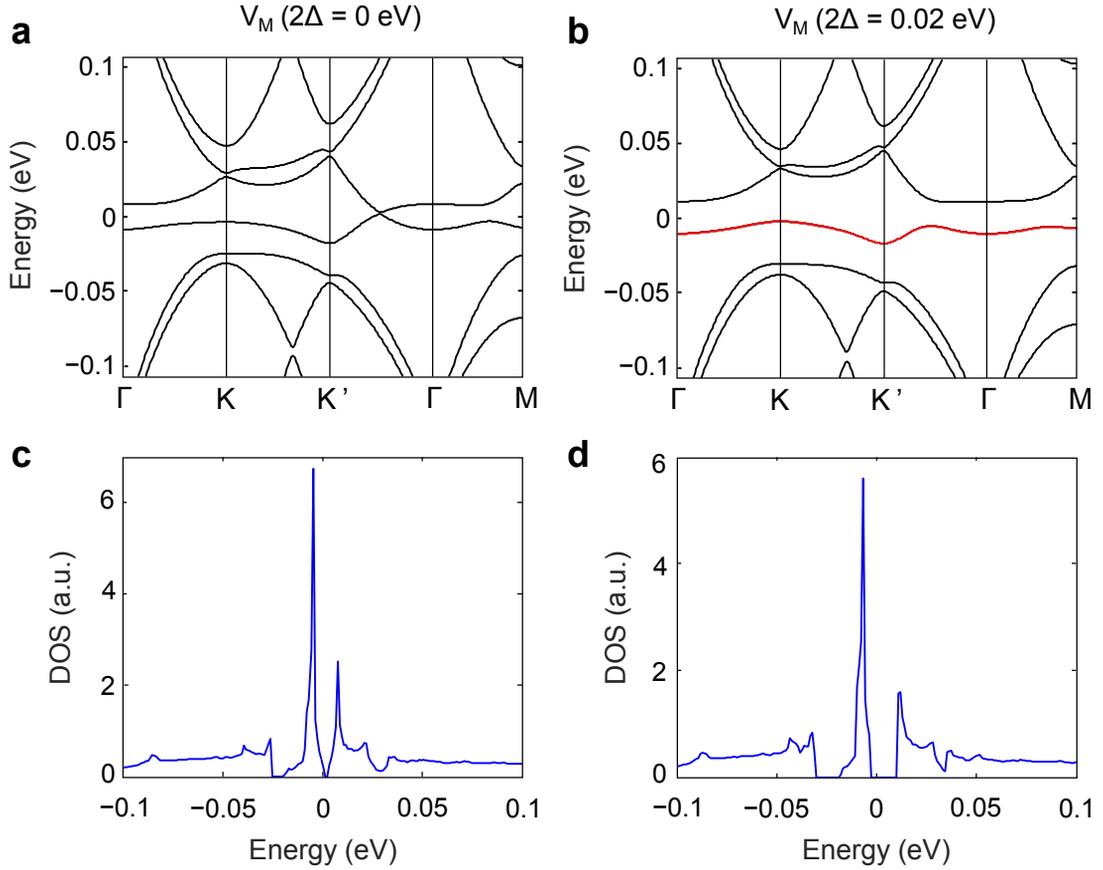

**Figure 4: Single-particle band structure of ABC TLG/hBN Moiré superlattice.
a and b**, Energy dispersion of the two electron and hole minibands without and with a vertical electrical field, respectively. The vertical electrical field in **b** generates a potential difference of 20 meV between the top and the bottom graphene layer. It leads to an isolated hole miniband with strongly suppressed bandwidth. The reduced electronic bandwidth relative to the Coulomb interaction enhances the electron correlation, and leads to the tunable Mott insulator states observed experimentally. **c** and **d**, Density of states (DOS) associated to the band structures in **a** and **b** that confirm the electron-hole asymmetry observed in our measurements. The band gap at charge neutrality due to the vertical electric field isolates the four-fold degenerate superlattice flat band of the hole side.